\documentclass[%
 aip,
 amsmath,amssymb,
 reprint,%
]{revtex4-1}

\usepackage{graphicx}% Include figure files
\usepackage{dcolumn}% Align table columns on decimal point
\usepackage{bm}% bold math
\usepackage[utf8]{inputenc}
\usepackage[T1]{fontenc}
\usepackage{mathptmx}
\usepackage{etoolbox}

\usepackage[colorlinks=true,
            linkcolor=blue,
            citecolor=blue,
            urlcolor=blue]{hyperref}

\makeatletter
\def\@email#1#2{%
 \endgroup
 \patchcmd{\titleblock@produce}
  {\frontmatter@RRAPformat}
  {\frontmatter@RRAPformat{\produce@RRAP{*#1\href{mailto:#2}{#2}}}\frontmatter@RRAPformat}
  {}{}
}%
\makeatother
\begin{document}

\preprint{AIP/123-QED}
\title{Magnetization-Tunable Topological Phase Transitions in Ferromagnetic Kagome Monolayers of Co$_3$X$_3$Y$_2$ ($X=\mathrm{Sn},\mathrm{Pb}$; $Y=\mathrm{S},\mathrm{Se}$)}

\author{Ritwik Das}
\email{intrd@iacs.res.in}
\affiliation{School of Physical Sciences, Indian Association for the Cultivation of Science \\
 2A and 2B Raja S.C. Mullick Road, Jadavpur, Kolkata 700 032, India}
\author{Arkamitra Sen}%
\affiliation{School of Physical Sciences, Indian Association for the Cultivation of Science \\
 2A and 2B Raja S.C. Mullick Road, Jadavpur, Kolkata 700 032, India}
\author{Indra Dasgupta}
\email{sspid@iacs.res.in}
\affiliation{School of Physical Sciences, Indian Association for the Cultivation of Science \\
 2A and 2B Raja S.C. Mullick Road, Jadavpur, Kolkata 700 032, India}

\date{\today}

\begin{abstract}
The quantum anomalous Hall effect in magnetic kagome materials has emerged as a versatile platform for dissipationless electronic and spintronic devices. In this work, we demonstrate that the orientation of magnetic moments $\hat{m}(\theta,\phi)$ at lattice sites provides a practical tuning mechanism for engineering nontrivial topological phases in monolayer kagome ferromagnets. To elucidate the mechanism, we construct a symmetry-adapted minimal tight-binding model for kagome ferromagnets that includes intrinsic spin-orbit coupling (SOC) and the intrinsic Rashba SOC permitted by broken out-of-plane mirror symmetry between nearest-neighbor kagome sites and can capture the resulting topological phase diagram as a function of $\hat{m}(\theta,\phi)$. In particular, the restoration of in-plane mirror symmetry for specific values of $\phi$ drives a topological phase transition upon varying the in-plane orientation of the moments $\hat{m}(\theta = 90^{\circ}, \phi)$. In contrast, for fixed $\phi$, the transitions driven by varying $\theta$ originate from the competition between Rashba SOC and intrinsic SOC. Density functional theory calculations for ferromagnetic kagome monolayers belonging to the Co$_3$X$_3$Y$_2$ family ($X=\mathrm{Sn},\mathrm{Pb}$; $Y=\mathrm{S},\mathrm{Se}$) support the predictions of the proposed minimal tight-binding model. These findings provide design guidelines for tunable topological phases in kagome materials.
\end{abstract}

\maketitle

%\begin{quotation}
%The ``lead paragraph'' is encapsulated with the \LaTeX\ 
%\verb+quotation+ environment and is formatted as a single paragraph before the first section heading. 
%(The \verb+quotation+ environment reverts to its usual meaning after the first sectioning command.) 
%Note that numbered references are allowed in the lead paragraph.

%The lead paragraph will only be found in an article being prepared for the journal \textit{Chaos}.
%\end{quotation}

Two-dimensional (2D) insulating monolayers with magnetic order offer a promising platform for exploring topological phases arising from the interplay of lattice geometry and spin-orbit coupling (SOC) \cite{Haldane, 2021_Anomalous_Hall_AFMs, 2022_Topological_aspects_of_AFMs, 2024_TQM_Kagome}. Insulating kagome ferromagnets, in particular, are intriguing due to their potential to host the quantum anomalous Hall effect (QAHE), characterized by a non-zero Chern number \cite{2010_Topological_insulators_RMP, ref_2, 2022_Topological_spintronics, 2022_Topological_kagome_magnets_and_superconductors} with dissipationless conducting edge states, making them viable candidates for applications in spintronics and quantum computing \cite{2000_Spin_anisotropy_and_QHE_in_kagome, 2011_QAHE_in_kagome}.

Recent studies on honeycomb materials have demonstrated that both out-of-plane and in-plane ferromagnetism can induce Chern insulating phases, resulting in topological phase diagrams that are tunable through the orientation of magnetic moments \cite{2011_QAHI_2D_electron_gas, 2016_QAHE_atomic_layers_inPlane_magnetization, 2017_OsCl3, 2019_2D_RT_FM_QAHE, 2022_Chern_number_tunable_QAHE, 2022_OsCl3_APL, 2023_H_FeCl2_APL, 2024_OsCl3}. Although minimal tight-binding (TB) models for such honeycomb systems suggest that in-plane ferromagnetism can generate nontrivial Chern insulating phases when both out-of-plane and in-plane mirror symmetries are broken, analogous phases and the associated topological phase transitions (TPT) in kagome ferromagnets with in-plane magnetization have received little attention \cite{2023_kagome_tune,2024_TQM_Kagome, 2025_review_hexagonal_2D}. The breaking and preservation of specific mirror symmetries are critical for TPT in these systems.

In this work, we develop a nearest-neighbor TB model that incorporates broken out-of-plane mirror symmetry to capture Chern insulating phases for ferromagnetic kagome systems where SOC terms are introduced using semiclassical arguments. This model successfully reproduces key topological properties, including TPT driven by variations of the orientation of in-plane magnetic moments $(\hat{m}(\theta=90^{\circ},\phi))$. While Chern insulators with in-plane moments in honeycomb systems typically exhibit Chern numbers \(C = \pm 1\), kagome systems can support higher Chern numbers, such as \(C = \pm 2\) for specific bands. Additionally, the model also predicts TPT for the out-of-plane orientation of magnetic moments $\hat{m}(\theta,\phi=\text{constant})$, where the bands may have a larger Chern numbers \(C = \pm 3\). Using first-principles electronic structure calculations within density functional theory (DFT), we demonstrate that the topological features predicted by this minimal TB model can be realised in insulating ferromagnetic monolayer family Co$_3$X$_3$Y$_2$ (X=Sn, Pb; Y=S, Se) while their bulk counterpart, specifically the ferromagnetic Weyl semimetal Co$_3$Sn$_2$S$_2$ also show magnetization tunable topological properties \cite{2019_bulk_Weyl, 2024_bulk_Weyl}. Importantly, recent experiments have demonstrated that the anomalous Hall response in magnetic materials can be systematically tuned by rotating the magnetization direction, highlighting the experimental feasibility of magnetization-controlled topological transport \cite{2019_AHE_angle_APL, 2024_MR_1, 2024_MR_2}.

\begin{figure}
\includegraphics{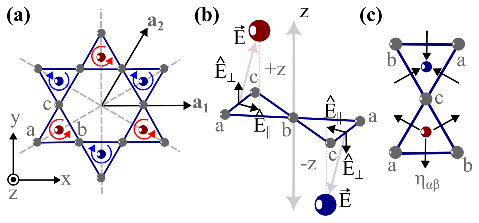}% Here is how to import EPS art
\caption{\label{kagome_structure} Kagome lattice geometry. \textbf{(a)} The kagome lattice with in-plane mirrors (grey dotted lines). Red (blue) arrows represent anticlockwise (clockwise) hoppings within the associated triangles, indicating the favored paths for up (down)-spins due to intrinsic SOC. The surrounding atoms above (red) and below (blue) the kagome plane break the out-of-plane mirror symmetry. \textbf{(b)} Schematic of the origin of first-nearest-neighbor SOC effect. \textbf{(c)} Directional convention used for $\vec{\eta}_{\alpha\beta}$ in the intrinsic Rashba SOC term of the TB model.}
\end{figure}

We begin by identifying the symmetries of the kagome lattice that determine the relevant terms allowed in the minimal TB model. Figure~\ref{kagome_structure}(a) shows the kagome lattice in a structural environment where the out-of-plane mirror symmetry, defined by a mirror plane residing at the kagome layer, is broken due to an asymmetric arrangement of surrounding atoms above and below the kagome plane. The in-plane mirror symmetries, indicated by the grey dotted lines in Figure~\ref{kagome_structure}(a), are preserved by the crystal structure in the absence of magnetic ordering and are perpendicular to the kagome plane. This simplified setting is motivated by the crystal structure of monolayer Co$_3$X$_3$Y$_2$ compounds, where analogous symmetry breaking arises from the local coordination environment.

The band dispersion due to the kagome geometry in the TB model is incorporated by including first-nearest-neighbor hopping terms ($t$), generating three spin-degenerate bands. An on-site field term ($B$) is added to account for ferromagnetism, thereby lifting the degeneracy and resulting in a six-band manifold.

SOC terms are introduced in the model semiclassically where the out-of-plane surrounding atoms provide an intrinsic electric field $\vec{E}$ at the centers of kagome bonds, which can be decomposed into two components: one parallel to the kagome plane ($\vec{E}_\parallel$) and another perpendicular to it ($\vec{E}_\perp$) (see Fig.~\ref{kagome_structure}(b)). When an electron hops between nearest-neighbor kagome sites, its velocity $\vec{v}$ couples with $\vec{E}$ to produce an effective magnetic field $\vec{B}_{\text{eff}} \propto \vec{v} \times \vec{E}$, which interacts with the electron's spin $\vec{\sigma}$, leading to an effective SOC Hamiltonian $\hat{H}_{SOC} = -\vec{\sigma} \cdot \vec{B}_{\text{eff}}$ \cite{Semisoc, 2024_OsCl3}. The SOC term from $\vec{E}_\parallel$ represents intrinsic SOC (I-SOC), while the term from $\vec{E}_\perp$ corresponds to intrinsic Rashba SOC (R-SOC). For I-SOC, the effective magnetic field $\vec{B}_{\text{eff}}$ is perpendicular to the kagome plane, while for R-SOC, a set of $\vec{\eta}_{\alpha\beta}$ vectors, parallel to the kagome plane and associated with each nearest-neighbor bond (see Fig.~\ref{kagome_structure}(c)), defines the direction of $\vec{B}_{\text{eff}}$. The R-SOC term survives only when the out-of-plane mirror symmetry $\hat{M}_z:(x,y,z)\rightarrow(x,y,-z)$ is broken by the surrounding atoms (see Fig.~\ref{kagome_structure}(b))\cite{2024_OsCl3}, which yields a finite out-of-plane electric field component $\vec{E}_\perp$.

Considering all these terms, we construct the following TB model:

\begin{align}
\Hat{H} =
& -t \sum_{\langle i\alpha\gamma, j\beta\gamma \rangle} c_{i\alpha\gamma}^{\dagger} c_{j\beta\gamma}
+ B \sum_{i\alpha\gamma\delta} c_{i\alpha\gamma}^{\dagger} (\hat{m}(\theta,\phi) \cdot \vec{\sigma})_{\gamma\delta} c_{i\alpha\delta} \notag \\
& + it_I \sum_{\langle i\alpha\gamma, j\beta\delta \rangle} \mu_{\alpha\beta} c_{i\alpha\gamma}^{\dagger} (\sigma_z)_{\gamma\delta} c_{j\beta\delta} \notag \\
& + i t_{R} \sum_{\langle i\alpha\gamma, j\beta\delta \rangle}
\nu_{\alpha\beta} \, c_{i\alpha\gamma}^{\dagger} \,(\vec{\sigma} \cdot \vec{\eta}_{\alpha\beta})_{\gamma\delta} \, c_{j\beta\delta} \label{eq}
\end{align}

Here, $\alpha, \beta$ are basis indices of the kagome lattice ($|a\rangle$, $|b\rangle$ and $|c\rangle$), and $\gamma, \delta$ represent the spin indices ($|\uparrow\rangle$ and $|\downarrow\rangle$). The second term corresponds to ferromagnetism, where $B$ represents the magnitude and the unit vector $\hat{m}(\theta,\phi)$ is the direction of the magnetic moment. The components of $\vec{\sigma}$ are the three Pauli matrices. The third and fourth terms correspond to the Kane-Mele type I-SOC \cite{2005_QSHE_Graphene} and intrinsic R-SOC, respectively. Importantly, unlike the second-nearest-neighbor SOC terms in the Haldane \cite{Haldane} and Kane-Mele \cite{2005_QSHE_Graphene} models for the honeycomb lattice \cite{2014_AHE_from_noncollinear_antiferromagnetism, 2019_Topological_states_breathing_kagome, 2022_magnetic_ordering_soc_kagome, ref_1, 2025_Ti3Se3X2_APL, 2025_Mn2XSe4_APL}, the SOC terms in the kagome lattice arise at first-nearest-neighbor level, as allowed by its local crystal symmetry (see Fig.~\ref{kagome_structure}(b)). In honeycomb systems the electric field at the midpoint of a nearest-neighbor bond vanishes by symmetry, whereas the asymmetric coordination in kagome systems permits a finite field resulting in first-nearest-neighbor SOC terms. The coefficients $\mu_{\alpha\beta}$ and $\nu_{\alpha\beta}$ encode the sign change under reversal of the hopping velocity $\vec{v}$, which reflects the opposite effective magnetic fields $\vec{B}_{\text{eff}}$ generated by clockwise and anticlockwise paths; therefore they take values of $\pm 1$, with the direction of hopping defined relative to the adjacent triangle formed by the nearest-neighbor bond under consideration (see Fig.~\ref{kagome_structure}(a)).

In the absence of R-SOC, i.e., $t_R = 0$, and with easy-axis ferromagnetism where the moments are aligned perpendicular to the kagome plane, the Hamiltonian in Eq.~\eqref{eq} breaks time-reversal symmetry $\hat{\Theta}$ while preserving inversion $\hat{I}$ and the out-of-plane mirror symmetry $\hat{M}_z$. As a consequence, the composite symmetries $\hat{\Theta}\otimes\hat{M}_z$, which makes the Berry curvature $\vec{\Omega}(\vec{k})$ odd within the first Brillouin zone (BZ) and $\hat{\Theta}\otimes\hat{M}_z\otimes\hat{I}$, which forces $\vec{\Omega}(\vec{k})$ to vanish at each $k$-point, are now broken, thereby allowing a nonzero Chern number $C$ \cite{2016_QAHE_atomic_layers_inPlane_magnetization}. In contrast, for in-plane moment alignment $\hat{m}(\theta=90^\circ,\phi)$ with $t_R = 0$, the Hamiltonian breaks both $\hat{\Theta}$ and $\hat{M}_z$ due to the pseudovector nature of the magnetic moment and the composite symmetries $\hat{\Theta}\otimes\hat{M}_z$ and $\hat{\Theta}\otimes\hat{M}_z\otimes\hat{I}$ remain preserved. As a consequence, the Berry curvature vanishes, and the I-SOC term alone cannot produce a finite Chern number for in-plane magnetization. The R-SOC term preserves both $\hat{\Theta}$ and $\hat{I}$, while breaking $\hat{M}_z$ irrespective of the direction of moments, thereby introducing non-trivial topology in the TB model for both in-plane and out-of-plane orientation of moments.

\begin{figure*}[ht] 
\centering
\includegraphics[width=1\textwidth]{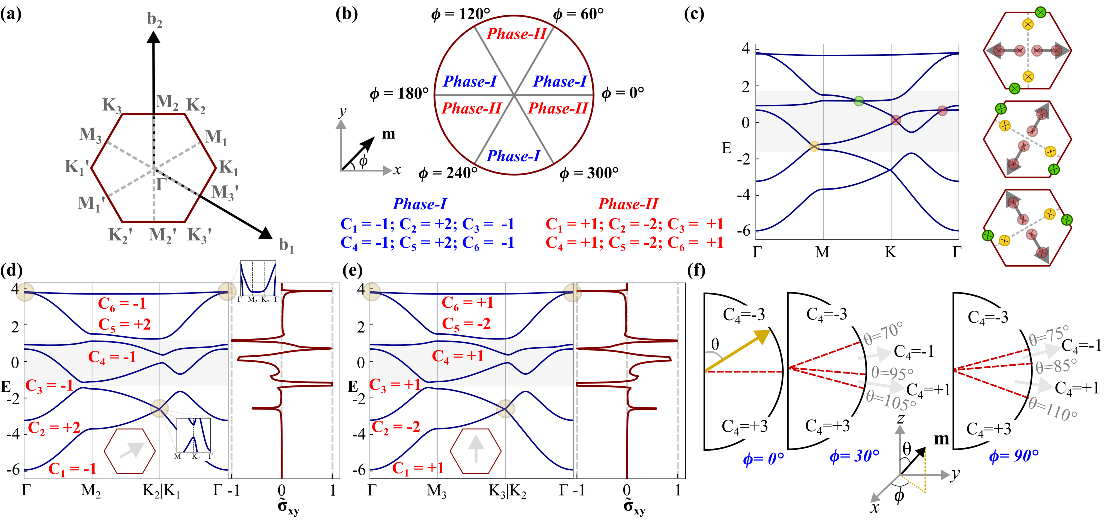} 
\caption{\label{tb_results} Topological properties of the TB model with parameters $t = 1.0$, $B = 1.5t$, $t_I = 0.2t$, and $t_R = 0.5t$. \textbf{(a)} First BZ of the kagome lattice with in-plane mirrors (grey dotted lines) and distinct high-symmetry points. \textbf{(b)} Topological phase diagram showing the Chern number $C$ for different bands as a function of $\hat{m}(\theta=90^{\circ}, \phi)$. \textbf{(c)} Band-crossing points due to ferromagnetic moment alignments $\hat{m}(\theta=90^\circ, \phi)$ for $\phi \in \{\phi_C\}$ indicating TPT. \textbf{(d)} Band structure and Chern number $C$ for each band at $\hat{m}(\theta=90^\circ, \phi=30^\circ)$. \textbf{(e)} Same as \textbf{(d)}, but for $\hat{m}(\theta=90^\circ, \phi=90^\circ)$. \textbf{(f)} TPT of the fourth band due to the variation of $\hat{m}(\theta, \phi=\text{constant})$ for $\phi = 0^\circ, 30^\circ$ and $90^\circ$.} 
\end{figure*}

The different topological phases of the TB model for $\hat{m}(\theta,\phi)$ are summarized in Fig.~\ref{tb_results}. The first BZ  with high-symmetry points $M_i$ ($M_1, M_2, M_3$) and $K_i$ ($K_1, K_2, K_3$) is shown in Fig.~\ref{tb_results}(a). All TB model parameters are chosen relative to \( t \). The strength of the ferromagnetic term \( B \leq 3.0t \) ensures that the spin-up and spin-down bands significantly mix due to SOC terms. The strength of \( t_R \) is chosen to dominate over \( t_I \) in order to realize nontrivial topological phases arising from the in-plane component of the moments.

As argued from symmetry considerations, the R-SOC induces Chern insulating phases for $\hat{m}(\theta=90^\circ,\phi)$. The topological phase diagram for variation of $\phi$ is shown in Fig.~\ref{tb_results}(b). The TPT is signalled by the closing of gaps at specific in-plane orientation of moments, such as $\phi = 0^\circ, 60^\circ, 120^\circ$ and their inverted counterparts with respect to the origin at $\phi = 180^\circ, 240^\circ, 300^\circ$ (this set of angles are denoted as $\{\phi_C\}$). The gap closes at $\{\phi_C\}$ as the direction of the in-plane moments are perpendicular to the in-plane mirrors thereby protecting the in-plane mirror symmetry for these orientations signaling TPT \cite{2013_InPlane_MagnetizationInduced_QAHE}. As the orientation of the in-plane moments traverses through the above symmetry protected phases, the Chern number C of each band flips, giving rise to two distinct phases, phase-$I$ and phase-$II$ as shown in Fig.~\ref{tb_results}(b) where the isolated kagome bands are numbered from $1$ to $6$ from bottom to top. The Chern numbers of the isolated bands are calculated using a standard Berry-curvature formalism, as detailed in the Supplementary Material \cite{Berry}.

\begin{figure*}[ht]
\centering
\includegraphics[width=1\textwidth]{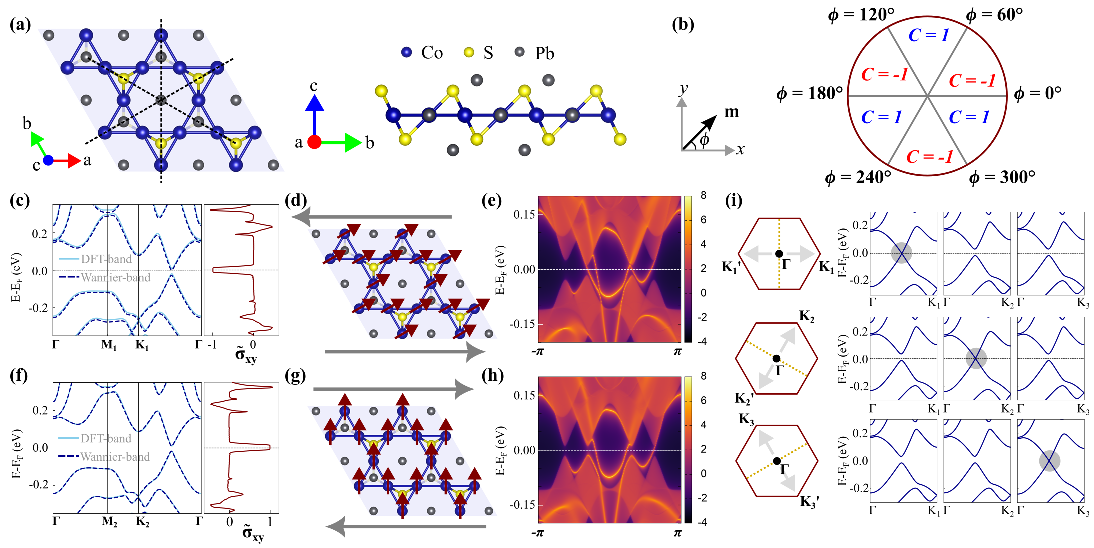}
\caption{\label{DFT_in_plane} Topological properties of the ferromagnetic Co$_3$Pb$_3$S$_2$ monolayer under in-plane moment variation. \textbf{(a)} Crystal structure showing in-plane mirrors (black dotted lines) and broken out-of-plane mirror symmetry (light-blue plane, coinciding with the kagome plane). \textbf{(b)} Topological phase diagram for in-plane moment variation ($\theta = 90^\circ$). \textbf{(c)} Electronic band structure for $\hat{m}(\theta = 90^{\circ}, \phi = 30^{\circ})$, with a Chern number $C = -1$ at $E_F$, as supported from the variation of the normalized Hall conductivity $\tilde{\sigma}_{xy}$. \textbf{(d)} Edge states for $\hat{m}(\theta = 90^{\circ}, \phi = 30^{\circ})$, with edge cuts perpendicular to the crystallographic axis $\vec{b}$. \textbf{(e)} Top-edge spectral function confirming $C = -1$ for the same edge cut. \textbf{(f)}–\textbf{(h)} Corresponding results for $\hat{m}(\theta = 90^{\circ}, \phi = 90^{\circ})$ with $C = +1$. \textbf{(i)} TPT for in-plane moment variation ($\theta = 90^\circ$), occurring cyclically at $\phi = 0^\circ$, $60^\circ$, and $120^\circ$, and their equivalents at $\phi = 180^\circ$, $240^\circ$, and $300^\circ$, tied to different $\Gamma$–$K$ paths parallel to the moment directions.}
\end{figure*}

We now focus on the band structure due to in-plane moment variations $(\hat{m}(\theta=90^{\circ}, \phi))$. When $\phi\in\{\phi_C\}$ the protected mirror symmetry forces all bands to touch, resulting in the absence of band gap, as shown in Fig.~\ref{tb_results}(c). The band crossing points (Weyl points in the $(k_x,k_y,\phi)$ parameter space for $\theta=90^{\circ}$) vary depending on the specific in-plane mirror symmetry being preserved. For example, for $\phi = 0^\circ$, band touching occurs along the $\Gamma-M_2-K_2$ and $K_1-\Gamma$ paths; while for $\phi = 60^\circ$, it occurs along $\Gamma-M_3-K_3$ and $K_2-\Gamma$ paths. All possible phase transition points between different bands with different in-plane moment orientations are also shown in Fig.~\ref{tb_results}(c). The variation of in-plane moments results in a non-trivial Chern number for each isolated band, as shown in Fig.~\ref{tb_results}(d) and (e) for two different moment orientations of $\hat{m}(\theta=90^{\circ},\phi=30^{\circ})$ and $\hat{m}(\theta=90^{\circ},\phi=90^{\circ})$ respectively where the changes in Chern numbers between them occur due to the TPT at $\phi=60^{\circ}$. It should be noted that the Chern number of each band can be defined as all the six bands for in-plane orientation of moments are isolated from each other and the apparent Weyl like degeneracies highlighted with yellow circles are indeed gapped (see inset of Fig.\ref{tb_results}(d)). The normalized Hall conductivity (\( \tilde{\sigma}_{xy}=\sigma_{xy}\frac{h}{e^2} \)) as a function of energy (see Fig. \ref{tb_results}(d) and (e)) as expected is an integer when it encounters a band gap.

TPTs also arise when the moment is tilted out of the plane, i.e., for $\hat{m}(\theta\neq 90^{\circ},\phi)$. These transitions, as we argue below, are driven by the competition between I-SOC and R-SOC, captured by the distinct parameters $t_I$ and $t_R$ in the TB model. In the limit $t_R=0$, varying $\theta$ at fixed $\phi$ produces a single transition at $\theta=90^\circ$, consistent with the aforementioned symmetry analysis and with the Onsager relation, which dictates that the Chern number changes sign under the reversal of moment \cite{2021_Anomalous_Hall_AFMs, 1984_landau}. When $t_R\neq 0$, the in-plane component of the moment ($\theta=90^\circ$) couples through $t_R$, while the out-of-plane component ($\theta=0^\circ$) primarily couples through $t_I$, and their competition generates multiple TPTs as the moment is rotated from $\theta=0^\circ$ to $180^\circ$, as illustrated for the fourth kagome band in Fig.~\ref{tb_results}(f). Here we focus on the fourth kagome band which carries the largest Chern number ($C_4=\pm 3$) and remains well isolated from the rest of the bands for the chosen parameter set and therefore admits an unambiguous assignment of Chern number for all orientations of $\hat{m}(\theta,\phi)$ \cite{isolate_chern}, except at the TPT points. Unlike in-plane orientation of moments, the other kagome bands do not remain isolated for variations of $\theta$ at constant $\phi$, and the Chern numbers—as well as the precise locations of TPTs—vary for different choices of TB parameters. Thus the minimal model is, in principle, capable of generating a wide variety of topological phase diagrams depending on parameter sets and on which gapped manifold of bands is considered (see Supplementary Material). Nevertheless, these variations do not alter the fundamental origin of the transitions, which arises from the restoration of an in-plane mirror symmetry for $\hat{m}(\theta=90^\circ,\phi)$ and from the competition between intrinsic and Rashba SOC when the moment is rotated out of the plane. Our analysis also reveals a simple symmetry of the phase diagram: if a TPT occurs at $(\theta,\phi)$, a corresponding transition appears at $(180^\circ-\theta,\phi+60^\circ)$. Using this relation together with the Onsager constraint, the full topological phase diagram of the TB model can be systematically constructed.

We apply the insights obtained from the minimal TB model to the ferromagnetic kagome monolayer Co$_3$Pb$_3$S$_2$ (space group $P\bar{3}m1$), a representative member of the Co$_3$X$_3$Y$_2$ family. Additional DFT results for other members and computational details are provided in the Supplementary Material \cite{2020_Chiral_edge_Co3Sn3S2, 2021_kagome_family, 2024_TPT_kagome, DFT_1, DFT_2, DFT_3, wan90_1, wan90_2, wantools}. The kagome structure of Co$_3$Pb$_3$S$_2$, featuring in-plane mirrors and broken out-of-plane mirror symmetry, is shown in Fig.~\ref{DFT_in_plane}(a). While the magnetic ground state is an easy-axis ferromagnet with moment $\approx 0.4\mu_B$ per Co, the magnetic anisotropy energy between the easy-axis and in-plane ferromagnetic orientation (along the $x$-axis) is 1.3 meV, suggesting that the moment orientations are experimentally accessible using external fields \cite{2022_exp}. 

Interestingly, the top most valence band for the kagome monolayer Co$_3$Pb$_3$S$_2$ that remain isolated from the lower manifold of bands, only contributes to the total Chern number, as the sum of the Chern numbers of the other occupied bands is zero irrespective of the orientation of the moments. In this sense, the topmost valence band plays a role analogous to the fourth band of the TB model as demonstrated for the chosen TB parameters. The topological phase diagram for varying in-plane orientation of moments in Co$_3$Pb$_3$S$_2$ is presented in Fig.~\ref{DFT_in_plane}(b), showing that the Chern number $C$ calculated for bands till $E_F$ varies between $-1$ and $+1$, consistent with the results obtained from the TB model (see Fig.~\ref{tb_results}(b)).

For an in-plane moment orientation at $\hat{m}(\theta=90^{\circ}, \phi=30^{\circ})$, detailed results are shown in Fig.~\ref{DFT_in_plane}(c) to (e). The band-gap is calculated to be 85 meV. The variation of $\tilde{\sigma}_{xy}$ near $E_F$ is shown in Fig.~\ref{DFT_in_plane}(c). Finite edge cuts (Fig.~\ref{DFT_in_plane}(d)) yield the surface spectrum of the top edge (Fig.~\ref{DFT_in_plane}(e)), supporting the $C=-1$ phase. Similarly results for $C=+1$, when the in-plane moment is along $\hat{m}(\theta=90^{\circ}, \phi=90^{\circ})$ is shown in Fig.~\ref{DFT_in_plane}(f) to (h). The TPT due to in-plane moment orientations occur for $\phi\in\{\phi_C\}$  when the moments are perpendicular to the in-plane mirrors, as in the TB model. At these points, conduction and valence bands touch at points along specific $\Gamma$–$K$ paths, which are perpendicular to the respective in-plane mirror planes where mirror symmetry is preserved \cite{2024_OsCl3}, as shown in Fig.~\ref{DFT_in_plane}(i).

\begin{figure*}
\centering
\includegraphics[width=0.8\textwidth]{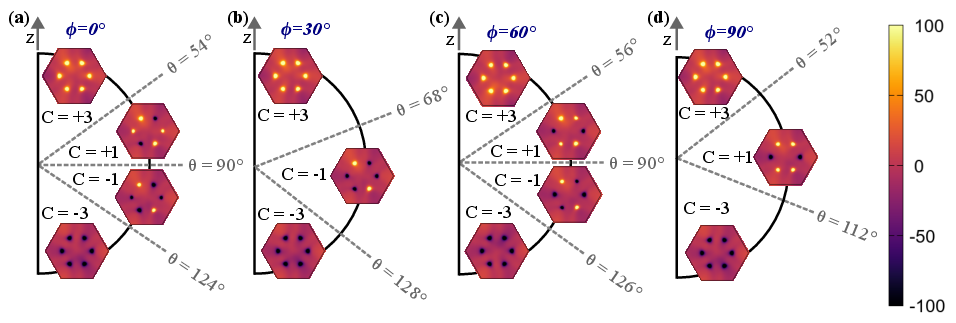}
\caption{\label{DFT_out_of_plane} TPT and Berry curvature of the top valence band within the first BZ for the ferromagnetic Co$_3$Pb$_3$S$_2$ monolayer under variation of moment $\hat{m}(\theta, \phi=$constant) for \textbf{(a)} $\phi = 0^\circ$, \textbf{(b)} $\phi = 30^\circ$, \textbf{(c)} $\phi = 60^\circ$, and \textbf{(d)} $\phi = 90^\circ$.}
\end{figure*}

The ground state of the monolayer Co$_3$Pb$_3$S$_2$ is an easy-axis ferromagnet with a Chern number $C = +3$ \cite{2021_kagome_family}, and our calculation reveal that it exhibit TPT upon the variation of the polar angle of the magnetic moment $\hat{m}(\theta,\phi=$constant). These transitions, shown for selected $\phi$ values in Fig.~\ref{DFT_out_of_plane}, closely resemble the results for the isolated fourth band of the TB model (Fig.~\ref{tb_results}(f)). The sign of the Chern numbers are however not in agreement between the DFT calculations and the prediction of the minimal TB model for the fourth band with the chosen set of parameters and may be attributed either to the chosen sign of the parameter $B$ in the TB model or the material specific details for the kagome monolayers \cite{2021_kagome_family}. 

The locations of the gap-closing points for $\hat{m}(\theta, \phi$=constant) are linked to the Berry curvature $\Omega_z(\vec{k})$, which exhibits peaks along all $\Gamma$–$K$ paths in the first BZ \cite{2024_TPT_kagome} (see Fig.~\ref{DFT_out_of_plane}). In the ground state $\hat{m}(\theta = 0^{\circ})$, six positive peaks in $\Omega_z(\vec{k})$ reflect the Chern number $C = +3$. Due to the variation of $\hat{m}(\theta,\phi=0^{\circ})$, the first TPT occurs at $\theta = 54^\circ$ along the $K_2$–$\Gamma$–$K_2'$ path, signaled by the reversal of $\Omega_z(\vec{k})$ peaks along this path. Since the TPT at $54^{\circ}$ is characterized by two Weyl points, therefore C decreases by 2, as shown in Fig.~\ref{DFT_out_of_plane}(a). Similar argument holds for all of the TPT shown in Fig.~\ref{DFT_out_of_plane}(a) to (d) \cite{2024_TPT_kagome} for $\phi=0^{\circ}, 30^{\circ}, 60^{\circ}$ and $90^{\circ}$ respectively.

The critical $\theta=\theta_C$ for the TPT for $\hat{m}(\theta \neq 90^\circ, \phi$=constant) is dependent on $t_R/t_I$ and therefore material dependent. As a consequence $\theta_C$ is no longer in agreement with the minimal TB model (Fig.~\ref{tb_results}(f)). This is further corroborated by calculating $\theta_C$ for other members of the same family (see Supplimentary Material). These quantitative variations do not, however, modify the mechanism of the transitions, which remains the interplay of the restorations of mirror symmetry and the competition between intrinsic and Rashba SOC. Interestingly, all the TPT points obey the relation $(\theta, \phi) \to (180^\circ - \theta, \phi + 60^\circ)$ (see Fig.~\ref{DFT_out_of_plane}), confirming the robustness of the minimal TB model.

In summary, we demonstrate that in-plane ferromagnetic Chern insulating phases and the associated topological phase transitions can also be realized in kagome materials, analogous to honeycomb systems \cite{2024_OsCl3}, where intrinsic Rashba SOC arising from broken out-of-plane mirror symmetry plays a crucial role for the topological behavior. Importantly, in contrast to honeycomb lattices, where SOC terms typically arise at the second-nearest-neighbor level, the kagome systems studied here host both intrinsic and Rashba SOC at the first-nearest-neighbor level due to electric fields generated by surrounding atoms. This leads to a distinct microscopic origin of the topological phases, which can be tuned by the orientation of the magnetic moments. Specific in-plane moment orientations $\hat{m}(\theta=90^{\circ},\phi\in\{\phi_C\})$ that preserve in-plane mirror symmetries give rise to a symmetry-governed topological phase structure for $\hat{m}(\theta=90^{\circ},\phi)$. Furthermore, we show that topological phase transitions driven by variations of $\hat{m}(\theta,\phi=\text{constant})$ originate from a competition between Rashba SOC and intrinsic SOC. In the absence of Rashba SOC, such transitions occur only at $\theta=90^{\circ}$, with the Chern number changing sign in accordance with symmetry constraints and the Onsager relation. The inclusion of Rashba SOC introduces additional features, resulting in a broader landscape of Chern insulating phases. We further propose a material realization in monolayer kagome compounds belonging to the Co$_3$X$_3$Y$_2$ family (X = Sn, Pb; Y = S, Se), which can potentially support the topological behavior captured by the minimal TB model. The resulting topological features, potentially accessible through external control of the magnetic moment, highlight kagome ferromagnets as promising platforms for tunable Berry-curvature and applications in quantum and spintronic technologies.

See the Supplementary Material for details of the computational method, additional tight-binding model results, and further topological properties of Co$_3$X$_3$Y$_2$ family materials ($X=\mathrm{Sn},\mathrm{Pb}$; $Y=\mathrm{S},\mathrm{Se}$) obtained from DFT calculations.

R.D thanks the Council of Scientific and Industrial Research (CSIR), India for research fellowship (File No. 09/080(1171)/2020-EMR-I). I.D would like to thank the Science and Engineering Research Board (SERB) India (Project No. CRG/2021/003024) and Technical Research Center, Department of Science and Technology Government of India for support.

\appendix

\begin{figure*}%[ht]
\centering
\includegraphics[width=0.9\linewidth]{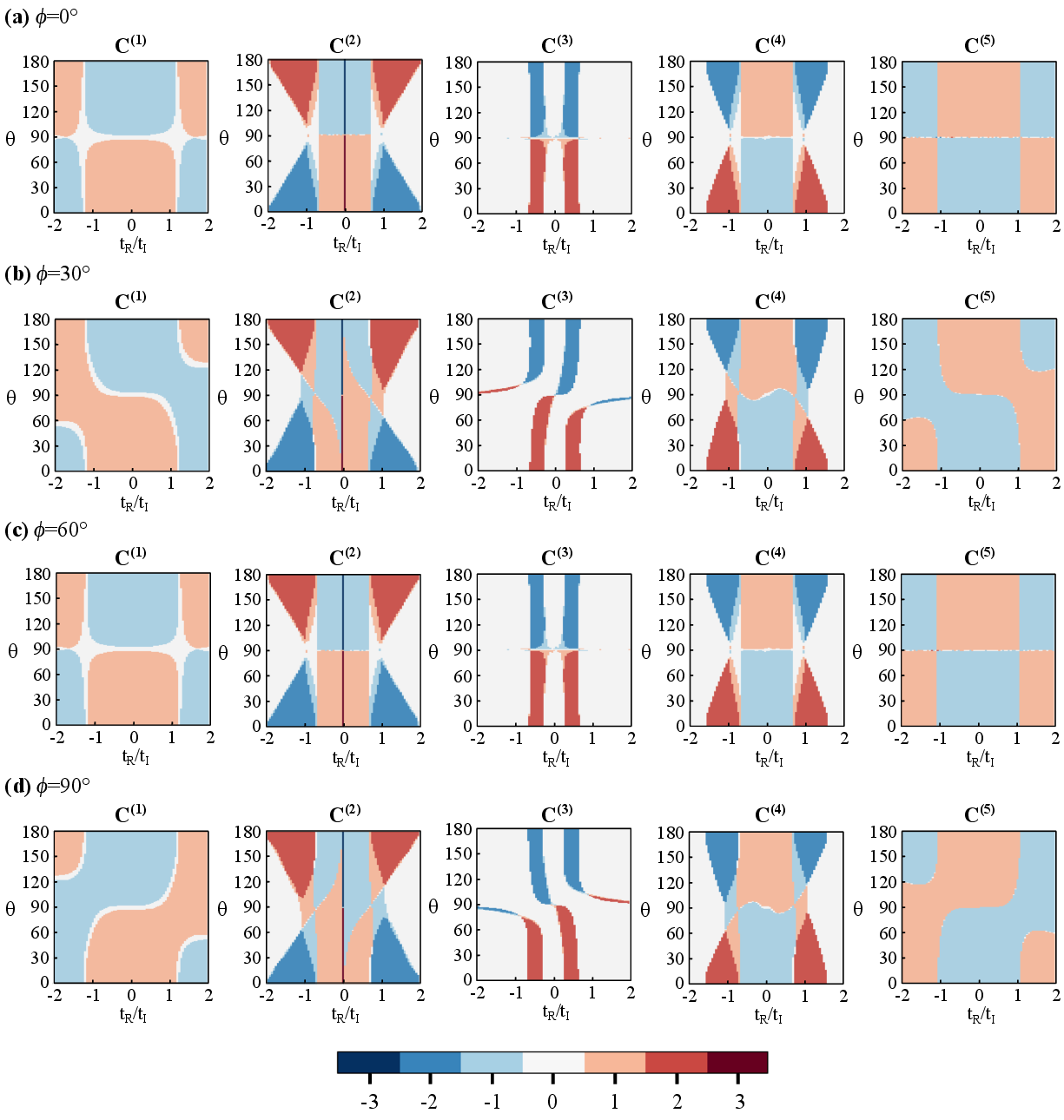}
\caption{\label{TB}Topological phase transitions of the TB model with parameters $t=1.0$ and $B=1.5t$ as a function of $t_R/t_I$ and the polar angle $\theta$ for fixed azimuthal angles $\phi=$ \textbf{(a)} $0^{\circ}$, \textbf{(b)} $30^{\circ}$, \textbf{(c)} $60^{\circ}$, and \textbf{(d)} $90^{\circ}$.}
\end{figure*}

\section{Computational Method}

DFT calculations were performed using the plane-wave-based projector augmented wave (PAW) method implemented in the Vienna \textit{ab initio} simulation package (VASP) with the generalized gradient approximation for exchange-correlation$^{41-43}$. The plane-wave cutoff was set to 500 eV, and a $\Gamma$-centered $12\times 12\times 1$ k-point mesh was used for BZ integration. Constrained-moment calculations fixed the Co-moment directions, while the Wannier90 and WannierTools codes were employed to construct the numerical TB-model retaining Co-d and Pb-p orbitals in the basis to analyse the topological properties of the chosen material$^{44-46}$.

To compute the Chern number of the $n$th isolated band for the tight binding (TB) model, we employ the standard Berry-curvature formalism$^{35}$,

\begin{align}
C_n = \frac{1}{2\pi}\int_{\vec{k} \in BZ} \Omega_n(\vec{k}) d^2k
\end{align}

where the Berry curvature is given by:

\begin{align}
\Omega_n(\vec{k}) = -\sum_{n' \neq n} \frac{2 \hbar^2 \, \text{Im}(\langle \psi_{n,\vec{k}}|\hat{v}_x|\psi_{n',\vec{k}}\rangle\langle \psi_{n',\vec{k}}|\hat{v}_y|\psi_{n,\vec{k}}\rangle)}{(E_n(\vec{k}) - E_{n'}(\vec{k}))^2}
\end{align}

where $E_n(\vec{k})$ and $\psi_{n,\vec{k}}$ are the energy and eigen-function of the electron in n-th band with crystal momentum $\vec{k}$ and $\hat{v}_{x(y)}$ is the velocity operator.

\begin{figure*}
\centering
\includegraphics[width=0.8\linewidth]{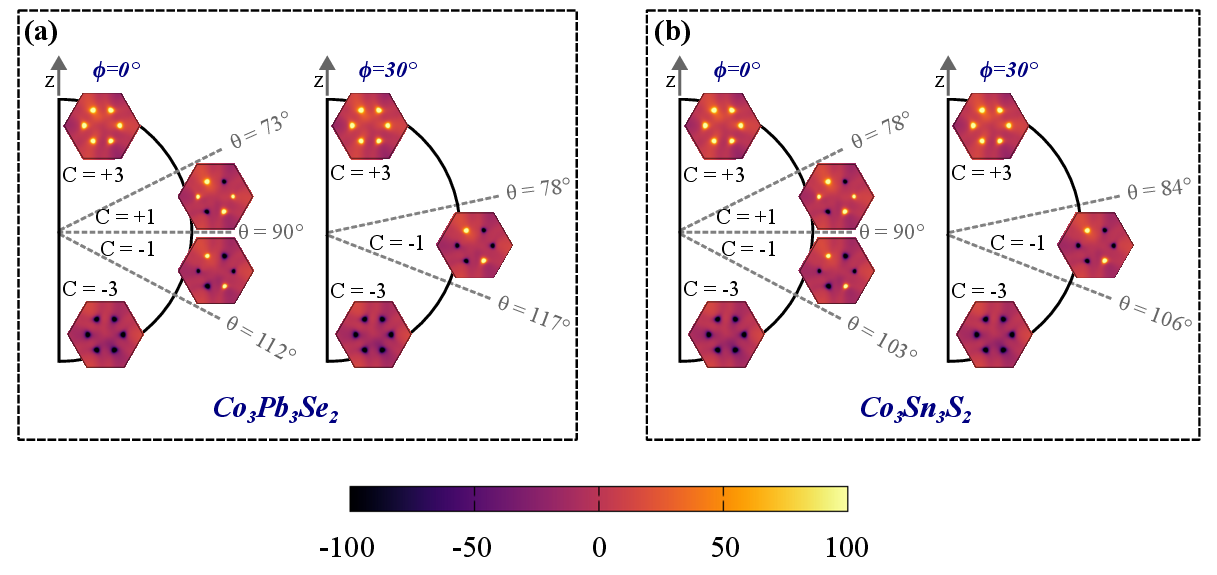}
\caption{\label{DFT} TPT and Berry curvature of the top valence band within the first BZ for the ferromagnetic Co$_3$X$_3$Y$_2$ monolayer under variation of moment $\hat{m}(\theta, \phi=$constant) for \textbf{(a)} Co$_3$Pb$_3$Se$_2$ and \textbf{(b)} Co$_3$Sn$_3$S$_2$}
\end{figure*}

\section{Topological properties of the TB model}

Fig.~\ref{TB} shows the topological phase transitions (TPTs) for different kagome-band fillings. We define
\begin{equation}
C^{(n)} = \sum_{m=1}^{n} C_m 
\end{equation}

where $C_m$ is the Chern number of the $m$-th band. Thus, $C^{(n)}$ denotes the total Chern number up to the $n$-th band filling. Most regions of the topological phase diagram in Fig.~\ref{TB} are characterized by integer-valued Chern numbers, except at the TPT regions. Fig.~\ref{TB} illustrates the variety of topological phases arising from the competition between intrinsic and Rashba spin-orbit coupling within the minimal TB model, where different band fillings or individual isolated bands can be considered to connect with the topological properties of real kagome ferromagnets with broken out-of-plane mirror symmetry.

Figure~\ref{TB} also highlights the influence of Rashba SOC ($t_R$) on TPTs for out-of-plane moment orientations $\hat{m}(\theta,\phi=\text{constant})$. We present results for $\phi=0^{\circ}$ [Fig.~\ref{TB}(a)] and $\phi=60^{\circ}$ [Fig.~\ref{TB}(c)], which belong to the set $\phi \in \{\phi_C\}$ and therefore exhibit TPTs at $\hat{m}(\theta=90^{\circ},\phi)$. In addition, results for $\phi=30^{\circ}$ [Fig.~\ref{TB}(b)] and $\phi=90^{\circ}$ [Fig.~\ref{TB}(d)] are shown. These results further support the general symmetry of the topological phase diagram, whereby a transition at $(\theta,\phi)$ is accompanied by a corresponding transition at $(180^{\circ}-\theta,\phi+60^{\circ})$.

\section{Additional Results for Co$_3$X$_3$Y$_2$ Family}

For in-plane orientations of moment, i.e., $\hat{m}(\theta=90^{\circ},\phi)$, the topological phase diagrams of Co$_3$Pb$_3$Se$_2$ and Co$_3$Sn$_3$S$_2$ are identical to that shown in Fig.~3(b) of the main text for Co$_3$Pb$_3$S$_2$. 
\noindent
Figures~\ref{DFT}(a) and \ref{DFT}(b) present the TPTs driven by tuning the out-of-plane orientation of moment, i.e., $\hat{m}(\theta,\phi=\text{constant})$, for $\phi=0^{\circ}$ and $\phi=30^{\circ}$ in Co$_3$Pb$_3$Se$_2$ and Co$_3$Sn$_3$S$_2$, respectively. Evidently, the critical angles for the TPTs differ among members of the monolayer Co$_3$X$_3$Y$_2$ family due to material-specific details. 

\nocite{*}
\bibliography{reference}% Produces the bibliography via BibTeX.

@article{2021_Anomalous_Hall_AFMs,
author={{\v{S}}mejkal, Libor and MacDonald, Allan H. and Sinova, Jairo and Nakatsuji, Satoru and Jungwirth, Tomas},
title={Anomalous Hall antiferromagnets},
journal={Nature Reviews Materials},
year={2022},
month={Jun},
day={01},
volume={7},
number={6},
pages={482-496},
issn={2058-8437},
doi={10.1038/s41578-022-00430-3},
url={https://doi.org/10.1038/s41578-022-00430-3}
}

@article{2022_Topological_aspects_of_AFMs,
doi = {10.1088/1361-6463/ac28fa},
url = {https://dx.doi.org/10.1088/1361-6463/ac28fa},
year = {2021},
month = {nov},
publisher = {IOP Publishing},
volume = {55},
number = {10},
pages = {103002},
author = {V Bonbien and Fengjun Zhuo and A Salimath and O Ly and A Abbout and A Manchon},
title = {Topological aspects of antiferromagnets},
journal = {Journal of Physics D: Applied Physics}
}

@article{2024_TQM_Kagome,
author={Wang, Qi and Lei, Hechang and Qi, Yanpeng and Felser, Claudia},
title={Topological Quantum Materials with Kagome Lattice},
journal={Accounts of Materials Research},
year={2024},
month={Jul},
day={26},
publisher={American Chemical Society},
volume={5},
number={7},
pages={786-796},
doi={10.1021/accountsmr.3c00291},
url={https://doi.org/10.1021/accountsmr.3c00291}
}

@article{2000_Spin_anisotropy_and_QHE_in_kagome,
  title = {Spin anisotropy and quantum Hall effect in the kagom\'e lattice: Chiral spin state based on a ferromagnet},
  author = {Ohgushi, Kenya and Murakami, Shuichi and Nagaosa, Naoto},
  journal = {Phys. Rev. B},
  volume = {62},
  issue = {10},
  pages = {R6065--R6068},
  numpages = {0},
  year = {2000},
  month = {Sep},
  publisher = {American Physical Society},
  doi = {10.1103/PhysRevB.62.R6065},
  url = {https://link.aps.org/doi/10.1103/PhysRevB.62.R6065}
}

@article{2011_QAHE_in_kagome,
doi = {10.1088/0953-8984/23/36/365801},
url = {https://dx.doi.org/10.1088/0953-8984/23/36/365801},
year = {2011},
month = {aug},
publisher = {IOP Publishing},
volume = {23},
number = {36},
pages = {365801},
author = {Zhi-Yong Zhang},
title = {The quantum anomalous Hall effect in kagomé lattices},
journal = {Journal of Physics: Condensed Matter}
}

@article{Berry,
  title = {Berry phase effects on electronic properties},
  author = {Xiao, Di and Chang, Ming-Che and Niu, Qian},
  journal = {Rev. Mod. Phys.},
  volume = {82},
  issue = {3},
  pages = {1959--2007},
  numpages = {0},
  year = {2010},
  month = {Jul},
  publisher = {American Physical Society},
  doi = {10.1103/RevModPhys.82.1959},
  url = {https://link.aps.org/doi/10.1103/RevModPhys.82.1959}
}

@article{2025_review_hexagonal_2D,
title = {The quantum anomalous Hall effect in two-dimensional hexagonal monolayers studied by first-principles calculations},
journal = {iScience},
volume = {28},
number = {1},
pages = {111622},
year = {2025},
issn = {2589-0042},
doi = {https://doi.org/10.1016/j.isci.2024.111622},
url = {https://www.sciencedirect.com/science/article/pii/S2589004224028499},
author = {Lixin Zhang and Hongxin Chen and Junfeng Ren and Xiaobo Yuan}
}

@article{ref_2,
  title = {Quantum anomalous Hall effect in graphene from Rashba and exchange effects},
  author = {Qiao, Zhenhua and Yang, Shengyuan A. and Feng, Wanxiang and Tse, Wang-Kong and Ding, Jun and Yao, Yugui and Wang, Jian and Niu, Qian},
  journal = {Phys. Rev. B},
  volume = {82},
  issue = {16},
  pages = {161414},
  numpages = {4},
  year = {2010},
  month = {Oct},
  publisher = {American Physical Society},
  doi = {10.1103/PhysRevB.82.161414},
  url = {https://link.aps.org/doi/10.1103/PhysRevB.82.161414}
}

@article{2023_kagome_tune,
  author  = {Lu, Jinlian and Xu, Xiaokang and Duan, Yuanyuan and Sun, Yi and Guan, Donghao and Chen, Anjie and Yao, Xiaojing and He, Ailei and Zhang, Xiuyun},
  title   = {Chern number transition of quantum anomalous hall phases in kagome {TM}$_3${T}e$_4$ ({TM} = {T}i, {C}r) monolayers by manipulating magnetization orientation},
  journal = {Applied Physics Letters},
  volume  = {123},
  number  = {13},
  pages   = {133102},
  year    = {2023},
  doi     = {10.1063/5.0164953}
}

@article{2022_Topological_kagome_magnets_and_superconductors,
author={Yin, Jia-Xin and Lian, Biao and Hasan, M. Zahid},
title={Topological kagome magnets and superconductors},
journal={Nature},
year={2022},
month={Dec},
day={01},
volume={612},
number={7941},
pages={647-657},
issn={1476-4687},
doi={10.1038/s41586-022-05516-0},
url={https://doi.org/10.1038/s41586-022-05516-0}
}

@article{ref_1,
doi = {10.1088/2053-1583/acfe88},
url = {https://dx.doi.org/10.1088/2053-1583/acfe88},
year = {2023},
month = {oct},
publisher = {IOP Publishing},
volume = {11},
number = {1},
pages = {011001},
author = {Mojarro, M A and Ulloa, Sergio E},
title = {Strain-induced topological transitions and tilted Dirac cones in kagome lattices},
journal = {2D Materials}
}

@article{2010_Topological_insulators_RMP,
  title = {Colloquium: Topological insulators},
  author = {Hasan, M. Z. and Kane, C. L.},
  journal = {Rev. Mod. Phys.},
  volume = {82},
  issue = {4},
  pages = {3045--3067},
  numpages = {0},
  year = {2010},
  month = {Nov},
  publisher = {American Physical Society},
  doi = {10.1103/RevModPhys.82.3045},
  url = {https://link.aps.org/doi/10.1103/RevModPhys.82.3045}
}

@article{2022_Topological_spintronics,
author={He, Qing Lin and Hughes, Taylor L. and Armitage, N. Peter and Tokura, Yoshinori and Wang, Kang L.},
title={Topological spintronics and magnetoelectronics},
journal={Nature Materials},
year={2022},
month={Jan},
day={01},
volume={21},
number={1},
pages={15-23},
issn={1476-4660},
doi={10.1038/s41563-021-01138-5},
url={https://doi.org/10.1038/s41563-021-01138-5}
}

@article{Haldane,
  title = {Model for a Quantum Hall Effect without Landau Levels: Condensed-Matter Realization of the "Parity Anomaly"},
  author = {Haldane, F. D. M.},
  journal = {Phys. Rev. Lett.},
  volume = {61},
  issue = {18},
  pages = {2015--2018},
  numpages = {0},
  year = {1988},
  month = {Oct},
  publisher = {American Physical Society},
  doi = {10.1103/PhysRevLett.61.2015},
  url = {https://link.aps.org/doi/10.1103/PhysRevLett.61.2015}
}

@article{2011_QAHI_2D_electron_gas,
  title = {Quantized anomalous Hall insulator in a nanopatterned two-dimensional electron gas},
  author = {Zhang, Yongping and Zhang, Chuanwei},
  journal = {Phys. Rev. B},
  volume = {84},
  issue = {8},
  pages = {085123},
  numpages = {5},
  year = {2011},
  month = {Aug},
  publisher = {American Physical Society},
  doi = {10.1103/PhysRevB.84.085123},
  url = {https://link.aps.org/doi/10.1103/PhysRevB.84.085123}
}

@article{2016_QAHE_atomic_layers_inPlane_magnetization,
  title = {Quantum anomalous Hall effect in atomic crystal layers from in-plane magnetization},
  author = {Ren, Yafei and Zeng, Junjie and Deng, Xinzhou and Yang, Fei and Pan, Hui and Qiao, Zhenhua},
  journal = {Phys. Rev. B},
  volume = {94},
  issue = {8},
  pages = {085411},
  numpages = {5},
  year = {2016},
  month = {Aug},
  publisher = {American Physical Society},
  doi = {10.1103/PhysRevB.94.085411},
  url = {https://link.aps.org/doi/10.1103/PhysRevB.94.085411}
}

@article{2017_OsCl3,
  title = {Monolayer of the $5d$ transition metal trichloride {O}s{C}l$_3$: A playground for two-dimensional magnetism, room-temperature quantum anomalous Hall effect, and topological phase transitions},
  author = {Sheng, Xian-Lei and Nikoli\ifmmode \acute{c}\else \'{c}\fi{}, Branislav K.},
  journal = {Phys. Rev. B},
  volume = {95},
  issue = {20},
  pages = {201402},
  numpages = {5},
  year = {2017},
  month = {May},
  publisher = {American Physical Society},
  doi = {10.1103/PhysRevB.95.201402},
  url = {https://link.aps.org/doi/10.1103/PhysRevB.95.201402}
}

@article{2019_2D_RT_FM_QAHE,
  title = {Two-Dimensional Room-Temperature Ferromagnetic Semiconductors with Quantum Anomalous Hall Effect},
  author = {You, Jing-Yang and Zhang, Zhen and Gu, Bo and Su, Gang},
  journal = {Phys. Rev. Appl.},
  volume = {12},
  issue = {2},
  pages = {024063},
  numpages = {7},
  year = {2019},
  month = {Aug},
  publisher = {American Physical Society},
  doi = {10.1103/PhysRevApplied.12.024063},
  url = {https://link.aps.org/doi/10.1103/PhysRevApplied.12.024063}
}

@article{2022_Chern_number_tunable_QAHE,
  title = {Chern Number Tunable Quantum Anomalous Hall Effect in Monolayer Transitional Metal Oxides via Manipulating Magnetization Orientation},
  author = {Li, Zeyu and Han, Yulei and Qiao, Zhenhua},
  journal = {Phys. Rev. Lett.},
  volume = {129},
  issue = {3},
  pages = {036801},
  numpages = {6},
  year = {2022},
  month = {Jul},
  publisher = {American Physical Society},
  doi = {10.1103/PhysRevLett.129.036801},
  url = {https://link.aps.org/doi/10.1103/PhysRevLett.129.036801}
}

@article{2013_InPlane_MagnetizationInduced_QAHE,
  title = {In-Plane Magnetization-Induced Quantum Anomalous Hall Effect},
  author = {Liu, Xin and Hsu, Hsiu-Chuan and Liu, Chao-Xing},
  journal = {Phys. Rev. Lett.},
  volume = {111},
  issue = {8},
  pages = {086802},
  numpages = {5},
  year = {2013},
  month = {Aug},
  publisher = {American Physical Society},
  doi = {10.1103/PhysRevLett.111.086802},
  url = {https://link.aps.org/doi/10.1103/PhysRevLett.111.086802}
}

@article{2019_bulk_Weyl,
  title = {Creating Weyl nodes and controlling their energy by magnetization rotation},
  author = {Ghimire, Madhav Prasad and Facio, Jorge I. and You, Jhih-Shih and Ye, Linda and Checkelsky, Joseph G. and Fang, Shiang and Kaxiras, Efthimios and Richter, Manuel and van den Brink, Jeroen},
  journal = {Phys. Rev. Res.},
  volume = {1},
  issue = {3},
  pages = {032044},
  numpages = {7},
  year = {2019},
  month = {Dec},
  publisher = {American Physical Society},
  doi = {10.1103/PhysRevResearch.1.032044},
  url = {https://link.aps.org/doi/10.1103/PhysRevResearch.1.032044}
}

@article{2024_bulk_Weyl,
  title = {Effective Model Analysis of Intrinsic Spin Hall Effect with Magnetism in the Stacked Kagome Weyl Semimetal {C}o$_3${S}n$_2${S}$_2$},
  author = {Ozawa, Akihiro and Kobayashi, Koji and Nomura, Kentaro},
  journal = {Phys. Rev. Appl.},
  volume = {21},
  issue = {1},
  pages = {014041},
  numpages = {9},
  year = {2024},
  month = {Jan},
  publisher = {American Physical Society},
  doi = {10.1103/PhysRevApplied.21.014041},
  url = {https://link.aps.org/doi/10.1103/PhysRevApplied.21.014041}
}

@book{2014_Khomskii, place={Cambridge}, title={Transition Metal Compounds}, publisher={Cambridge University Press}, author={Khomskii, Daniel I.}, year={2014}}

@article{Semisoc,
  title = {Low-energy effective Hamiltonian involving spin-orbit coupling in silicene and two-dimensional germanium and tin},
  author = {Liu, Cheng-Cheng and Jiang, Hua and Yao, Yugui},
  journal = {Phys. Rev. B},
  volume = {84},
  issue = {19},
  pages = {195430},
  numpages = {11},
  year = {2011},
  month = {Nov},
  publisher = {American Physical Society},
  doi = {10.1103/PhysRevB.84.195430},
  url = {https://link.aps.org/doi/10.1103/PhysRevB.84.195430}
}

@article{2024_OsCl3,
doi = {10.1088/2516-1075/ad4b81},
url = {https://dx.doi.org/10.1088/2516-1075/ad4b81},
year = {2024},
month = {may},
publisher = {IOP Publishing},
volume = {6},
number = {2},
pages = {025005},
author = {Ritwik Das and Subhadeep Bandyopadhyay and Indra Dasgupta},
title = {In-plane magnetization orientation driven topological phase transition in {O}s{C}l$_3$ monolayer},
journal = {Electronic Structure}
}

@article{2014_AHE_from_noncollinear_antiferromagnetism,
  title = {Anomalous Hall Effect Arising from Noncollinear Antiferromagnetism},
  author = {Chen, Hua and Niu, Qian and MacDonald, A. H.},
  journal = {Phys. Rev. Lett.},
  volume = {112},
  issue = {1},
  pages = {017205},
  numpages = {5},
  year = {2014},
  month = {Jan},
  publisher = {American Physical Society},
  doi = {10.1103/PhysRevLett.112.017205},
  url = {https://link.aps.org/doi/10.1103/PhysRevLett.112.017205}
}

@article{2024_MR_1,
  title = {Orbital Magneto-Nonlinear Anomalous Hall Effect in Kagome Magnet {F}e$_3${S}n$_2$},
  author = {Wang, Lujunyu and Zhu, Jiaojiao and Chen, Haiyun and Wang, Hui and Liu, Jinjin and Huang, Yue-Xin and Jiang, Bingyan and Zhao, Jiaji and Shi, Hengjie and Tian, Guang and Wang, Haoyu and Yao, Yugui and Yu, Dapeng and Wang, Zhiwei and Xiao, Cong and Yang, Shengyuan A. and Wu, Xiaosong},
  journal = {Phys. Rev. Lett.},
  volume = {132},
  issue = {10},
  pages = {106601},
  numpages = {7},
  year = {2024},
  month = {Mar},
  publisher = {American Physical Society},
  doi = {10.1103/PhysRevLett.132.106601},
  url = {https://link.aps.org/doi/10.1103/PhysRevLett.132.106601}
}

@article{2025_Ti3Se3X2_APL,
  author  = {Li, Jie and Xu, Xiaokang and Mao, Yuqing and Lu, Jinlian and Chang, Xinghao and Liu, Yuxuan and He, Ailei and Zhang, Xiuyun},
  title   = {{Two-dimensional triangular-lattice {Ti$_3$Se$_3$X$_2$} ({X} = {S}, {Te}) monolayer: Stable quantum anomalous {H}all insulator with high temperature and high {C}hern number}},
  journal = {Applied Physics Letters},
  volume  = {127},
  number  = {7},
  pages   = {073103},
  year    = {2025},
  doi     = {10.1063/5.0277667}
}

@article{2025_Mn2XSe4_APL,
  author  = {Yao, Xiaojing and Li, Jiahui and Li, Jie and Xu, Xiaokang and Wang, Zijin and He, Ailei and Lu, Jinlian and Zhang, Xiuyun},
  title   = {{Quantum anomalous {H}all effect in two-dimensional ferromagnetic {Mn$_2$XSe$_4$} ({X} = {Al}, {Ga}, {In})}},
  journal = {Applied Physics Letters},
  volume  = {126},
  number  = {22},
  pages   = {223101},
  year    = {2025},
  doi     = {10.1063/5.0264468}
}

@article{2024_MR_2,
  title = {In-Plane Anomalous Hall Effect Associated with Orbital Magnetization: Measurements of Low-Carrier Density Films of a Magnetic Weyl Semimetal},
  author = {Nakamura, Ayano and Nishihaya, Shinichi and Ishizuka, Hiroaki and Kriener, Markus and Watanabe, Yuto and Uchida, Masaki},
  journal = {Phys. Rev. Lett.},
  volume = {133},
  issue = {23},
  pages = {236602},
  numpages = {6},
  year = {2024},
  month = {Dec},
  publisher = {American Physical Society},
  doi = {10.1103/PhysRevLett.133.236602},
  url = {https://link.aps.org/doi/10.1103/PhysRevLett.133.236602}
}

@article{2019_Topological_states_breathing_kagome,
  title = {Topological states on the breathing kagome lattice},
  author = {Bolens, Adrien and Nagaosa, Naoto},
  journal = {Phys. Rev. B},
  volume = {99},
  issue = {16},
  pages = {165141},
  numpages = {7},
  year = {2019},
  month = {Apr},
  publisher = {American Physical Society},
  doi = {10.1103/PhysRevB.99.165141},
  url = {https://link.aps.org/doi/10.1103/PhysRevB.99.165141}
}

@article{2019_AHE_angle_APL,
  author  = {Chen, R. Y. and Zhang, R. Q. and Liao, L. Y. and Chen, X. Z. and Zhou, Y. J. and Gu, Y. D. and Saleem, M. S. and Zhou, X. F. and Pan, F. and Song, C.},
  title   = {{Magnetic field direction dependent magnetization reversal in synthetic antiferromagnets}},
  journal = {Applied Physics Letters},
  volume  = {115},
  number  = {13},
  pages   = {132403},
  year    = {2019},
  doi     = {10.1063/1.5118928}
}

@article{2022_magnetic_ordering_soc_kagome,
author = {Watanabe ,Jin and Araki ,Yasufumi and Kobayashi ,Koji and Ozawa ,Akihiro and Nomura ,Kentaro},
title = {Magnetic Orderings from Spin–Orbit Coupled Electrons on Kagome Lattice},
journal = {Journal of the Physical Society of Japan},
volume = {91},
number = {8},
pages = {083702},
year = {2022},
doi = {10.7566/JPSJ.91.083702},

URL = {https://doi.org/10.7566/JPSJ.91.083702},
eprint = {https://doi.org/10.7566/JPSJ.91.083702}
}

@article{2005_QSHE_Graphene,
  title = {Quantum Spin Hall Effect in Graphene},
  author = {Kane, C. L. and Mele, E. J.},
  journal = {Phys. Rev. Lett.},
  volume = {95},
  issue = {22},
  pages = {226801},
  numpages = {4},
  year = {2005},
  month = {Nov},
  publisher = {American Physical Society},
  doi = {10.1103/PhysRevLett.95.226801},
  url = {https://link.aps.org/doi/10.1103/PhysRevLett.95.226801}
}

@book{1984_landau,
  address = {New York},
  author = {Landau, L. D. and Lifshitz, E. M.},
  comment = {Mezzi Continui},
  publisher = {Pergamon},
  title = {Electrodynamics of Continuous Media},
  year = 1984
}

@article{2022_OsCl3_APL,
  author  = {Yu, Yawei and Xie, Xiao and Liu, Xiaobiao and Li, Jia and Peeters, Fran{\c c}ois M. and Li, Linyang},
  title   = {{Two-dimensional semimetal states in transition metal trichlorides: A first-principles study}},
  journal = {Applied Physics Letters},
  volume  = {121},
  number  = {11},
  pages   = {112405},
  year    = {2022},
  doi     = {10.1063/5.0105605}
}

@article{2020_Chiral_edge_Co3Sn3S2,
  title = {Emerging chiral edge states from the confinement of a magnetic Weyl semimetal in {C}o$_3${S}n$_2${S}$_2$},
  author = {Muechler, Lukas and Liu, Enke and Gayles, Jacob and Xu, Qiunan and Felser, Claudia and Sun, Yan},
  journal = {Phys. Rev. B},
  volume = {101},
  issue = {11},
  pages = {115106},
  numpages = {6},
  year = {2020},
  month = {Mar},
  publisher = {American Physical Society},
  doi = {10.1103/PhysRevB.101.115106},
  url = {https://link.aps.org/doi/10.1103/PhysRevB.101.115106}
}

@article{2023_H_FeCl2_APL,
  author  = {Yang, Xin and Shen, Yanqing and Lv, Lingling and Zhou, Min and Zhang, Yu and Meng, Xianghui and Jiang, Xiangqian and Ai, Qing and Shuai, Yong and Zhou, Zhongxiang},
  title   = {{Tuning the topological phase and anomalous Hall conductivity with magnetization direction in {H-FeCl$_2$} monolayer}},
  journal = {Applied Physics Letters},
  volume  = {123},
  number  = {20},
  pages   = {203102},
  year    = {2023},
  doi     = {10.1063/5.0175382}
}

@article{2024_TPT_kagome,
  title = {Topological transitions by magnetization rotation in kagome monolayers of the ferromagnetic Weyl semimetal Co-based shandite},
  author = {Nakazawa, Kazuki and Kato, Yasuyuki and Motome, Yukitoshi},
  journal = {Phys. Rev. B},
  volume = {110},
  issue = {8},
  pages = {085112},
  numpages = {10},
  year = {2024},
  month = {Aug},
  publisher = {American Physical Society},
  doi = {10.1103/PhysRevB.110.085112},
  url = {https://link.aps.org/doi/10.1103/PhysRevB.110.085112}
}

@article{2021_kagome_family,
  title = {Kagome quantum anomalous Hall effect with high Chern number and large band gap},
  author = {Zhang, Zhen and You, Jing-Yang and Ma, Xing-Yu and Gu, Bo and Su, Gang},
  journal = {Phys. Rev. B},
  volume = {103},
  issue = {1},
  pages = {014410},
  numpages = {8},
  year = {2021},
  month = {Jan},
  publisher = {American Physical Society},
  doi = {10.1103/PhysRevB.103.014410},
  url = {https://link.aps.org/doi/10.1103/PhysRevB.103.014410}
}

@article{DFT_1,
  title = {Projector augmented-wave method},
  author = {Bl\"ochl, P. E.},
  journal = {Phys. Rev. B},
  volume = {50},
  issue = {24},
  pages = {17953--17979},
  numpages = {0},
  year = {1994},
  month = {Dec},
  publisher = {American Physical Society},
  doi = {10.1103/PhysRevB.50.17953},
  url = {https://link.aps.org/doi/10.1103/PhysRevB.50.17953}
}

@article{DFT_2,
  title = {Ab initio molecular dynamics for liquid metals},
  author = {Kresse, G. and Hafner, J.},
  journal = {Phys. Rev. B},
  volume = {47},
  issue = {1},
  pages = {558--561},
  numpages = {0},
  year = {1993},
  month = {Jan},
  publisher = {American Physical Society},
  doi = {10.1103/PhysRevB.47.558},
  url = {https://link.aps.org/doi/10.1103/PhysRevB.47.558}
}

@article{DFT_3,
  title = {Generalized Gradient Approximation Made Simple},
  author = {Perdew, John P. and Burke, Kieron and Ernzerhof, Matthias},
  journal = {Phys. Rev. Lett.},
  volume = {77},
  issue = {18},
  pages = {3865--3868},
  numpages = {0},
  year = {1996},
  month = {Oct},
  publisher = {American Physical Society},
  doi = {10.1103/PhysRevLett.77.3865},
  url = {https://link.aps.org/doi/10.1103/PhysRevLett.77.3865}
}

@article{wan90_1,
doi = {10.1088/1361-648X/ab51ff},
url = {https://dx.doi.org/10.1088/1361-648X/ab51ff},
year = {2020},
month = {jan},
publisher = {IOP Publishing},
volume = {32},
number = {16},
pages = {165902},
author = {Giovanni Pizzi and Valerio Vitale and Ryotaro Arita and Stefan Blügel and Frank Freimuth and Guillaume Géranton and Marco Gibertini and Dominik Gresch and Charles Johnson and Takashi Koretsune and Julen Ibañez-Azpiroz and Hyungjun Lee and Jae-Mo Lihm and Daniel Marchand and Antimo Marrazzo and Yuriy Mokrousov and Jamal I Mustafa and Yoshiro Nohara and Yusuke Nomura and Lorenzo Paulatto and Samuel Poncé and Thomas Ponweiser and Junfeng Qiao and Florian Thöle and Stepan S Tsirkin and Małgorzata Wierzbowska and Nicola Marzari and David Vanderbilt and Ivo Souza and Arash A Mostofi and Jonathan R Yates},
title = {Wannier90 as a community code: new features and applications},
journal = {Journal of Physics: Condensed Matter}
}

@article{Wan90_2,
  title = {Maximally localized Wannier functions: Theory and applications},
  author = {Marzari, Nicola and Mostofi, Arash A. and Yates, Jonathan R. and Souza, Ivo and Vanderbilt, David},
  journal = {Rev. Mod. Phys.},
  volume = {84},
  issue = {4},
  pages = {1419--1475},
  numpages = {0},
  year = {2012},
  month = {Oct},
  publisher = {American Physical Society},
  doi = {10.1103/RevModPhys.84.1419},
  url = {https://link.aps.org/doi/10.1103/RevModPhys.84.1419}
}

@article{wantools,
title = {WannierTools: An open-source software package for novel topological materials},
journal = {Computer Physics Communications},
volume = {224},
pages = {405-416},
year = {2018},
issn = {0010-4655},
doi = {https://doi.org/10.1016/j.cpc.2017.09.033},
url = {https://www.sciencedirect.com/science/article/pii/S0010465517303442},
author = {QuanSheng Wu and ShengNan Zhang and Hai-Feng Song and Matthias Troyer and Alexey A. Soluyanov}
}

@article{2022_exp,
  title = {Tuning the electronic band structure in a kagome ferromagnetic metal via magnetization},
  author = {Kumar, Neeraj and Soh, Y. and Wang, Yihao and Li, Junbo and Xiong, Y.},
  journal = {Phys. Rev. B},
  volume = {106},
  issue = {4},
  pages = {045120},
  numpages = {6},
  year = {2022},
  month = {Jul},
  publisher = {American Physical Society},
  doi = {10.1103/PhysRevB.106.045120},
  url = {https://link.aps.org/doi/10.1103/PhysRevB.106.045120}
}

@article{isolate_chern,
  title = {Flat Chern Band in a Two-Dimensional Organometallic Framework},
  author = {Liu, Zheng and Wang, Zheng-Fei and Mei, Jia-Wei and Wu, Yong-Shi and Liu, Feng},
  journal = {Phys. Rev. Lett.},
  volume = {110},
  issue = {10},
  pages = {106804},
  numpages = {5},
  year = {2013},
  month = {Mar},
  publisher = {American Physical Society},
  doi = {10.1103/PhysRevLett.110.106804},
  url = {https://link.aps.org/doi/10.1103/PhysRevLett.110.106804}
}

\end{document}